# Laser-Stabilized Fiber Sensing on Terrestrial Field-Deployed Cable for Enhanced Sensitivity

**Rajiv Boddeda[1], Adrish Sahu[1], Christian Dorize[1], Arnaud Dupas[1], Pierre Brochard[2], Haïk Mardoyan[1], Carina Castineiras[1] and Jérémie Renaudier[1]**

*(1) Nokia Bell Labs, 12 rue Jean Bart, 91300 Massy, France (2) Silentsys, 10 rue Xavier Bichat, 72000, Le Mans, France*

rajiv.boddeda@nokia-bell-labs.com

**Abstract:** We demonstrate distributed acoustic sensing on deployed buried cable leveraging laser frequency stabilization and probing code optimization. Our results confirm the capability to accurately localize weak events with high resolution and sensitivity over 80 km. 

## 1. Introduction

Reusing existing optical telecom infrastructure for distributed environmental sensing enables large-scale, cost-efficient monitoring without deploying dedicated sensing fibers [1,2]. Conventional distributed fiber-optic sensing (DFS) techniques still face practical limits that hinder scalable deployment: sensitivity and sensing range degrade with distance [3]. In practice, polarization fading, laser frequency noise, and the persistent trade-off between spatial resolution and optical reach are dominant technical bottlenecks. To overcome this, coherent multiple-input multiple-output (MIMO) DFS architectures were developed, where digitally coded phase modulation jointly probes two orthogonal polarization axes using a polarization-diversity coherent receiver [4,5]. This configuration inherently mitigates polarization fading and enables coexistence with wavelength-division multiplexing (WDM) data traffic [6,7]. This concept can be extended by introducing a laser stability compensation module that mitigates low-frequency (<1 MHz) fluctuations in the sensing laser source, which are shown through simulation to be a dominant source of DFS sensitivity degradation [8]. So far, most published field demonstrations have so far relied on either short [3], highly exposed fiber segments [7] or on large near-fiber disturbances to extract signals [2], providing limited evidence that DFS can robustly detect weak environmental signals over long, buried telecom links.

We present experimental validation of the proposed system when the fiber is deployed using a conventional buried approach under hard ground and in proximity to highways and other noise sources. We corroborate prior numerical analyses and employ a recently developed laser stabilization system to further enhance sensitivity. Applying frequency-noise compensation to our sensing setup operating over an 80 km standard single-mode fiber (SSMF) link—with the final ~20 km traversing mixed urban and rural environments—yields a pronounced sensitivity improvement, enabling detection of environmental vibrations within a 382 Hz mechanical bandwidth and at 1 m spatial resolution. These results highlight the potential of stabilized MIMO-DFS architectures for high-performance, long-reach sensing over buried telecom fiber cables.

## 2. Impact of probing code type and laser stabilization on DFS sensitivity

Fiber sensing operates by transmitting designed probing codes over a set duration, where the code length, frequency content, and type (e.g., Golay sequences or frequency sweeps) can be tailored. In Fig. 1(a), we show the MIMO-DFS setup where the probing codes are encoded on the two polarisations before being sent to the Fiber Under Test (FUT). Rayleigh backscattered light from the FUT is isolated via an optical circulator and directed to a coherent receiver, before performing digital signal processing (DSP) to recover the complex field to estimate phase and intensity as functions of distance. To test the long-range sensitivity of the setup, we use 60 km fiber spool in a lab environment connected to either 20 km of fiber spool or a deployed fiber buried under ground of approximately the same length. Fig 1(b), shows the FUT route extending through urban and rural environments, crossing highways and bridges. The inset presents the duct cable's transverse cross-section, showing an outer sheath of high-density polyethylene supported by thermoplastic rods. Six tubes house different fiber types, with the FUT located in the red tube. The durable duct fiber cable exhibits a reduced mechanical impulse response at the fiber, which—together with the material's tensile strength—significantly attenuates and complicates fiber sensing, particularly at high frequencies.

In Fig. 1(c), we show our compact experimental setup with the code generator and the transmitter with laser on the top. Most laser sources exhibit rising frequency noise at low Fourier frequencies around 10 Hz–10 kHz band, precisely where DFS sensitivity is most affected. Hence, we use an optical frequency discriminator (OFD) to convert laser frequency fluctuations to voltage via a reference arm. Compared to the previous OFD [8], the new OFD from Silentsys offers improved frequency noise floor at low Fourier frequencies thanks to a novel temperature control of the optical frequency discrimination module that keeps the internal fiber scheme at the micro-Kelvin scale for hours. The resulting signal drives a servo controller feeding the frequency-modulation input of the sensing dedicated Indie LXM U® laser source, as shown in Fig. 1(a).

Our probing employs advanced polarization multiplexed M-ary phase shift keying codes derived from [4], which allows us to achieve a shorter code length that approaches the physical limit of the FUT. The code is made of $2^{17}$ symbols that modulate the laser light with 1 dBm output power at 100 Mbaud, leading to a 1 m native gauge length, and the permanent repetition of the code allows to capture mechanical events over a 382 Hz bandwidth. Successful detection of the probing codes sent to the FUT enables estimation of a spatially resolved snapshot of the distributed scatterers' phase profile. As the backscattered signal traverses the fiber link twice, it suffers double-pass attenuation, which substantially lowers the signal level at long ranges as shown in Fig. 2(a). We also show the differential phase deviation ($\sigma_\varphi$) estimated as function of distance on well-isolated fiber spools in the lab (60+20 km) using only laser internal stabilization and a high pass filter of 20 Hz. We observe that it rises rapidly beyond 30 km and beyond 60 km it becomes difficult to resolve events with phase variations below 1 radian in 20 Hz. Hence, precise estimation of phase information from the fiber's far end requires stringent noise control in both the transmitter and receiver, while leveraging laser coherence and robust DSP techniques to enhance SNR and preserve phase fidelity across distance. Moreover, using longer codes enables more averaging time and thus more accurate phase extraction; however, it also increases the time between phase snapshots, which limits the highest acoustic frequencies that can be detected.

To further enhance sensitivity and spatial resolution, we optimized the interrogation code by increasing the symbol rate and increasing the frequency range. A higher symbol rate effectively reduces the gauge length, thereby improving spatial localization of perturbations, though it also lowers the optical power available per symbol. However, because Rayleigh backscattering is inherently sporadic along the fiber, the received signal can be optimized by selecting the strongest backscattering points through controlled down sampling. Figure 2(b) compares the differential phase variation as a function of distance for two different symbol rates, 25 Mbaud (red) and 100 Mbaud (green) by selecting the best scatterer over a fixed length of ~50 meters. Even with laser frequency stabilization, the standard deviation of phase rises beyond 60 km due to insufficient back reflected intensity over a symbol duration with 25 Mbaud. Increasing the symbol rate to 100 Mbaud significantly reduces the error in the differential phase estimation, confirming the advantage of higher-rate interrogation for higher sensitivity. This improvement, however, comes at the cost of increased data throughput. Nevertheless, recent developments in compressive sensing and data reduction algorithms provide efficient strategies to manage the resulting data volume without compromising measurement fidelity [9].

### 3. Experimental results

We quantified the distributed fiber sensing (DFS) phase standard deviation as a function of distance and converted it into the equivalent fiber elongation per meter following the calibration method [10]. We experimentally assessed this using the 80 km of SSMF lab spools and then compare with the one from the field. The resulting measured fiber strain is shown in green in Fig. 2(c), we remain below 30 nm/m for fiber lengths up to 80 km, demonstrating that the observed increase in phase variance originates primarily from laser frequency drift rather than optical power limitations. This confirms our system's ability to perform sensing across spans typical of standard single-mode fiber links in terrestrial deployments, where buried, bundled fiber severely attenuates the signal.

Now, we connect the deployed fiber (shown in grey in Fig. 1(c)) to our setup to estimate the fiber strain as a function of distance. To validate the setup's acoustic-sensing capability, we applied an acoustic excitation at the 62 km point along the link by placing an audio speaker near the duct cable and generating a 300 Hz tone at approximately 65 dBA. As illustrated by the blue curve in Fig. 2(c), the corresponding strain exhibits a clear and localized peak, despite the sensing fiber being well insulated in a multi-fiber bundle. The inset presents the power spectral density (PSD) of the phase time series at that location, where the 300 Hz tone is distinctly observed confirming the large audio bandwidth of our fiber sensing setup. Note that we observe the cable termination at ~ 82 km. Furthermore, we observe varying strain levels along the fiber, predominantly below 50 nm/m, despite the duct cable running near highways, bridges, and industrial areas. These low-level strain signals would likely have remained undetected with standard MIMO-DFS setup. Subsequently, we performed continuous wavelet analysis along the full link to identify and classify different environmental signatures as shown in Fig. 2 (d). Unlike controlled field trials, we observe a superposition of multiple events here, making isolation of individual signatures more challenging. Measurements were taken at various times of day and at night to capture differing activity levels. Even under low-disturbance conditions (at 2 a.m.), localized wavelet signatures appeared near 62.6 km, corresponding to intermittent traffic at a small suburban roundabout. At 72 km, a broader, time-varying spectral response was observed, attributed to continuous vehicle motion on a nearby highway. Finally, at 78 km, distinct bidirectional traffic patterns were identified on an overpass bridge, demonstrating the MIMO-DFS system's ability to discriminate multiple, spatially separated vibration sources. Future work will focus on building a database of reference templates representing known vibration signatures to support anomaly detection, enabling proactive cable condition monitoring and minimizing false alarm rates.

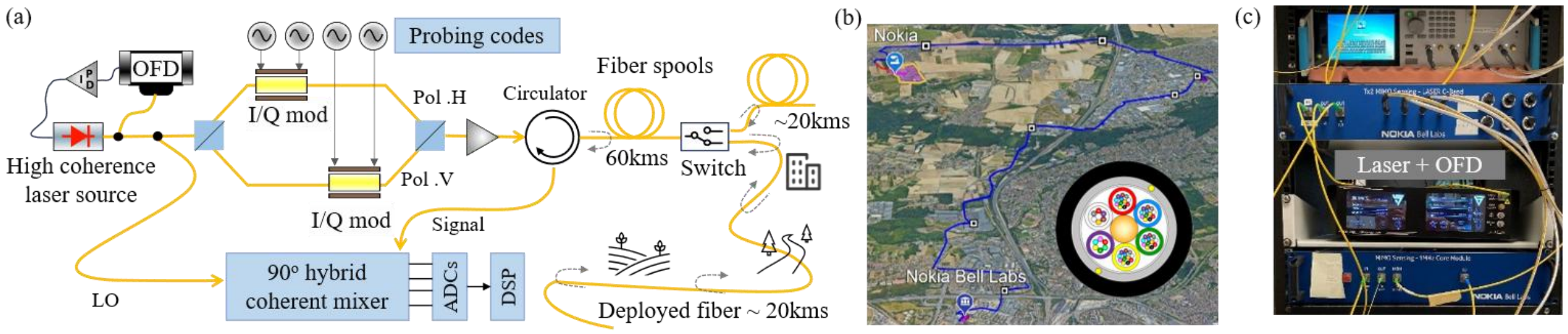

Fig. 1: (a) MIMO-DFS setup depicting the laser stabilization with the deployed fiber. (b) The deployed fiber crossing urban and rural environments, (inset) shows the transverse section of the fiber cable. (c). The experimental setup shows the key components—namely, the code generator with the transmitter and the OFD and the receiver at the bottom.

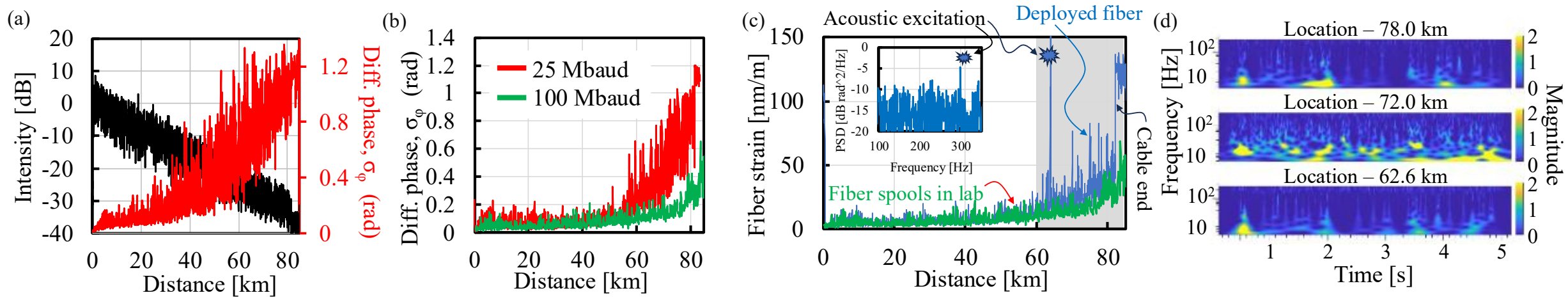

Fig. 2: a) The reflected intensity and the differential phase ($\sigma_\varphi$) as a function of distance measured over 80 kms of fiber spools in the lab using a commercial laser. b) The differential phase measurement using 25 Mbaud and 100 Mbaud codes, c) Measured Fiber strain along the 80 km of isolated SSMF in the lab (green) and the trace in blue is the hybrid 60 km (lab) and 20 km of deployed fiber (in grey) showing an acoustic excitation at 62 km and the inset shows the PSD showing the detected frequency of 300 Hz. (d) Scalogram of MIMO-DFS at different locations showing the detected acoustic signature of urban and rural environment at 2 AM.

## Conclusion

We enhanced laser stabilization and employed selective down sampling of dominant Rayleigh scatterers, thereby mitigating the per-symbol power deficit intrinsic to high-rate interrogation and improving phase estimation accuracy. Consequently, we achieved sensitivity below 30 nm/m over distances up to 80 km. After validating event detection, wavelet analysis revealed distinct, location-specific vibration signatures from human activity, confirming that high-rate, laser-stabilized MIMO-DFS enables robust distributed sensing over buried telecom fiber cables.

Acknowledgments: This work has been supported by the EU's Horizon Europe, under grant agreement No 101189545 – SENSEI